\documentclass[sigconf]{acmart}

\usepackage{CJKutf8}
\usepackage{graphicx}
\usepackage{array}

\usepackage[utf8]{inputenc}
\usepackage[super]{nth}
\usepackage{fancyhdr,amsmath,amsthm,subfig,url,longtable,placeins,tabto,enumitem,color,colortbl,geometry,etoolbox,multirow,makecell,pifont}
\usepackage{xspace}
\usepackage{afterpage}
\usepackage{soul}

\AtBeginDocument{%
  \providecommand\BibTeX{{%
    \normalfont B\kern-0.5em{\scshape i\kern-0.25em b}\kern-0.8em\TeX}}}

\setcopyright{acmlicensed}
\copyrightyear{2025}
\acmYear{2025}
\setcopyright{cc}
\setcctype{by}
\acmConference[CHI '25]{CHI Conference on Human Factors in Computing
Systems}{April 26-May 1, 2025}{Yokohama, Japan}
\acmBooktitle{CHI Conference on Human Factors in Computing Systems (CHI
'25), April 26-May 1, 2025, Yokohama,
Japan}
\acmDOI{10.1145/3706598.3713616}
\acmISBN{979-8-4007-1394-1/25/04}

\newif\ifstatus
\statustrue

\begin{document}
\begin{CJK*}{UTF8}{gbsn}

\title{Exploring the Privacy and Security Challenges Faced by Migrant Domestic Workers in Chinese Smart Homes}


\author{Shijing He}
\email{shijing.he@kcl.ac.uk}
\affiliation{
  \institution{King's College London}
  \city{London}
  \country{United Kingdom}
}

\author{Xiao Zhan}
\email{xiao.zhan@kcl.ac.uk}
\affiliation{
  \institution{King's College London}
  \city{London}
  \country{United Kingdom}
}

\author{Yaxiong Lei}
\email{yl212@st-andrews.ac.uk}
\affiliation{
  \institution{University of St Andrews}
  \city{St Andrews}
  \country{United Kingdom}
}

\author{Yueyan Liu}
\email{yueyanliu.hci@gmail.com}
\affiliation{
  \institution{China Academy of Art}
  \city{Hangzhou}
  \country{China}
}

\author{Ruba Abu-Salma}
\email{ruba.abu-salma@kcl.ac.uk}
\affiliation{
  \institution{King's College London}
  \city{London}
  \country{United Kingdom}
}

\author{Jose Such}
\email{jose.such@kcl.ac.uk}
\affiliation{
  \institution{King's College London}
  \city{London}
  \country{United Kingdom}
}








\renewcommand{\shortauthors}{He et al.}

\begin{abstract}
The growing use of smart home devices poses considerable privacy and security challenges, especially for individuals like migrant domestic workers (MDWs) who may be surveilled by their employers. This paper explores the privacy and security challenges experienced by MDWs in multi-user smart homes through in-depth semi-structured interviews with 26 MDWs and 5 staff members of agencies that recruit and/or train domestic workers in China. Our findings reveal that the relationships between MDWs, their employers, and agencies are characterized by significant power imbalances, influenced by Chinese cultural and social factors (such as Confucianism and collectivism), as well as legal ones. Furthermore, the widespread and normalized use of surveillance technologies in China, particularly in public spaces, exacerbates these power imbalances, reinforcing a sense of constant monitoring and control. Drawing on our findings, we provide recommendations to domestic worker agencies and policymakers to address the privacy and security challenges facing MDWs in Chinese smart homes.
\end{abstract}




\maketitle

\section{Introduction}
Homes around the world are getting smarter in recent years. The rapid adoption of smart home devices has significantly raised privacy and security concerns among various stakeholders, 
including device owners and primary users~\cite{zeng2017end,zheng2018user,abdi2019more,huang2020amazon,rodriguez2024you,le2024towards}, as well as bystanders~\cite{albayaydh2024co,despres2024my,zhou2024bring,park2024well,marky2024decide,chiang2024more,feger2023connectivitycontrol,shalawadi2024manual,wang2023exploring}. In particular, recent literature has explored the privacy and security needs, concerns, and preferences of at-risk populations, including bystander groups 
who 
have little to no access to these devices and face challenges like power imbalances and privacy invasions in their living and/or work environments, as seen with migrant domestic workers (MDWs) 
who work in smart homes owned by their employers~\cite{bernd2020bystanders,bernd2022balancing,slupska2022they,albayaydh2023examining,albayaydh2023innovative,albayaydh2022exploring,johnson2020beyond,ju2023re}. 
However, significantly fewer studies have examined this topic in the context of mainland China from a non-WEIRD (Western, Educated, Industrialized, Rich, Democratic) perspective~\cite{hasegawa2024weird}. 
As a non-WEIRD country, China presents distinct factors that make this investigation both timely and interesting, including trends in digitalization and the widespread adoption of smart home devices—more than 78 million Chinese households used smart home devices in 2023~\cite{chinadata}—as well as emerging laws (see \textbf{\S \ref{subsec:regulations}}). Furthermore, in 2021, China had 37.6 million MDWs, making up about 8\% of the total national employment. Of these MDWs, 90\% were women who migrated from rural to urban areas, and 60\% had high school education or lower. The domestic service industry included 2.617 million companies, the majority of which were small and medium-sized, accounting for 62\% of the total~\cite{ndrc}. Additionally, previous research on MDWs has not explored the challenges they face, such as power imbalances between MDWs and labor recruitment agencies, or their views and practices related to mitigating privacy concerns in both their workplaces and daily lives (e.g., using smart home cameras to monitor elderly family members). Studies have yet to explore how these dynamics affect MDWs, agencies, employers, and household members within China’s unique cultural and social context (e.g., Confucian values and collectivism), and to examine how gaps in legal protections impact vulnerable groups such as MDWs. 

In this paper, we focus on the substantial 90\% (approximately 34 million) of MDWs in China, who are individuals migrating from \emph{rural/developing} areas to \emph{urban/more developed} regions, usually seeking better opportunities for domestic work. These include: 1) live-in MDWs, who temporarily reside in their employers' homes for work-related reasons, most commonly as nannies, babysitters, and postnatal caregivers; 
and 2) live-out MDWs, who work in their employers' homes but do not reside there, typically staying for short periods, and are often house cleaners, housekeepers, maintenance workers, and porters. These MDWs often experience significant vulnerabilities, such as precarious recruitment, lack of privacy and safety protections, social and cultural isolation, limited access to employment information, poor enforcement of labor laws, and restrictions on their movement and association~\cite{ILO2023}. Furthermore, these MDWs have increasingly adopted smart home devices for family care and security purposes due to their dispersed locations, such as using surveillance cameras to monitor the safety of their property~\cite{h2022monitoring}. Chinese MDWs often rely on domestic worker agencies to match them with employers, aligning their skills with the specific needs of the household. These agencies offer services such as screening, placement, contract management, training, certification, and conflict resolution. Therefore, we also explore the crucial role these agencies play in assisting employers with recruiting, evaluating, and managing MDWs for both permanent and temporary household positions. 

Our paper aims to answer the following research questions (RQs):

\begin{itemize}
\item \textbf{RQ1.} What are the privacy and security concerns, views, and needs of Chinese MDWs, and how do domestic worker agencies address these issues through their policies and practices?

\item \textbf{RQ2.} How do the experiences of Chinese MDWs, including the privacy and security challenges they face, impact power dynamics and influence their relationships with agencies and employers in the workplace?

\item \textbf{RQ3.} How do Chinese MDWs negotiate and resolve privacy- and security-related conflicts with their family members or employers regarding the use of smart home devices?
\end{itemize}

To answer our RQs, we conducted 31 semi-structured interviews in China, including 26 with MDWs and 5 with staff members of domestic worker agencies. We examined the privacy and security concerns, needs, and perspectives of MDW participants, noting that they faced significant privacy issues, particularly with surveillance cameras, as bystanders in their employers' homes (see \textbf{\S\ref{awareness and perspectives}}). We also investigated the challenges faced by MDW participants, including power imbalances in employer-employee relationships and inadequate support from agencies (see \textbf{\S\ref{perspectives on PSS in workplace}}). Additionally, we aimed to understand the strategies employed by MDW participants to address privacy concerns both in their workplaces and within their own smart homes. We found that the lack of legal protections and the unclear regulatory landscape in China exacerbated the vulnerable position of MDW participants, such as the absence of regulations and the lack of defined laws governing the domestic service industry, highlighting the urgent need for updates to privacy laws and regulations to better protect MDW rights (see \textbf{\S\ref{mitigation practices and suggestions}}). We also investigated the privacy perspectives and practices of Chinese domestic worker agencies (see \textbf{\S\ref{perspectives on PSS in workplace}}, \textbf{\S \ref{mitigation practices and suggestions}}).

Our findings reveal that Chinese MDW participants frequently expressed concerns about personal information security, the excessive collection of personal data, and the intrusive nature of surveillance devices, especially cameras, in their workplaces. They often had limited access to privacy and security advice, instead relying on consultation hotlines, government channels, and word-of-mouth communication. Privacy concerns among MDW participants varied according to factors like gender, age, personality, and socioeconomic status, and were also shaped by cultural influences such as collectivism and the normalization of public surveillance. MDW participants also encountered privacy and security challenges, such as employer surveillance, limited support from agencies, and restricted advocacy for privacy rights. In response, our MDW participants developed coping strategies, including negotiating privacy measures with employers and avoiding sensitive conversations near surveillance devices. Although agencies recognized MDWs' privacy needs, they often prioritized employer demands, underscoring the need for regulatory changes to better safeguard the privacy and labor rights of MDWs in China.

Our paper makes the following contributions. First, unlike previous studies, our research is the first to highlight the influence of Chinese social, cultural, and legal norms on the privacy and security perceptions and practices of MDWs, emphasizing the need for transparency in surveillance practices within MDWs' workplaces (see \textbf{\S\ref{contributions}}). Second, to our knowledge, our study is the first to offer empirical insights into the social, cultural, and legal factors that shape Chinese MDWs' privacy and security perceptions and practices in multi-user smart home environments (see \textbf{\S\ref{suggestions to public}}). Lastly, this paper explores the power dynamics between MDWs, agencies, and employers, offering a more localized understanding of privacy challenges and providing practical recommendations to enhance legal protections for Chinese MDWs working in multi-user smart home environments (see \textbf{\S\ref{suggestions to agencies}}, \textbf{\S\ref{legal suggestion}}).

\section{Related Work} \label{related}
\subsection{Multi-User Smart Home Privacy and Security} \label{subsec:multiuser}
Recent work has explored the privacy and security concerns, needs, and preferences of smart home device owners/primary users~\cite{zheng2018user,haney2021s,abdi2019more,abdi2021privacy,yao2019defending,thakkar2022would,zeng2017end,barbosa2019if,tabassum2019idont,haney2023user,lau2018alexa,huang2020amazon,mare2020smart,li2023will,huang2019perception,levinson2024snitches,chalhoub2020factoring,chalhoub2020innovation}; overviews can be found in~\cite{pattnaik2023survey,lipford2022privacy}. These studies highlight the profound influence of social norms, power dynamics, and technical knowledge on perceptions and adoption of smart home devices. They underscore the importance of privacy and transparency in data practices (e.g., data collection, data deletion) as key factors in promoting broader acceptance and use of these technologies~\cite{lau2018alexa,huang2020amazon,abdi2021privacy,apthorpe2018discovering,chalhoub2021did,geeng2019s,song2020m,tabassum2019idont,colnago2020informing,he2018rethinking,sovacool2021knowledge,singh2018users,marky2020you,abbott2022privacy,emami2019exploring,zeng2017end,he2024contextualizing}. 
Furthermore, previous research has examined the privacy and security concerns and requirements of bystanders who may be affected by smart home devices owned by others, such as invited guests, visiting friends, passersby, rental tenants, neighbors, and roommates~\cite{yao2019privacy,alshehri2023exploring,cobb2021would,mare2020smart,geeng2019s,marky2020idont,h2022monitoring,pierce2022addressing,pierce2019smart,despres2024my,chiang2024more,rodriguez2019towards,tabassum2023exploring,albayaydh2024co,zhou2024bring,park2024well,marky2024decide,chiang2024more,feger2023connectivitycontrol,shalawadi2024manual,wang2023exploring,abdi2022home}, as well as other household members (including those living in hostile environments)~\cite{freed2019my,alshehri2020smart,leitao2019anticipating}. 

Previous research has also investigated the privacy and security needs and concerns of MDWs, highlighting challenges associated with resettlement, communication, and power imbalances in smart homes within WEIRD contexts~\cite{bernd2020bystanders,bernd2022balancing,slupska2022they,abu2024they}. For example, Ming et al. have examined how technologies such as electronic visit verification (EVV) systems, which monitored care workers without their consent, intensified power imbalances and strained employment relationships~\cite{ming2024wage}. However, limited research on multi-user smart home privacy has explored the perspectives of MDWs in non-WEIRD contexts. 
For example, Johnson et al. have investigated how Filipino MDWs in Hong Kong perceived and responded to workplace surveillance cameras, revealing that monitoring intended to ensure care often eroded trust and negatively impacted quality of care~\cite{johnson2020beyond}. Albayaydh et al. have examined the privacy concerns of MDWs in Jordan, highlighting the intricate interplay between religious and social norms, legal frameworks, and factors like gender
~\cite{albayaydh2022exploring,albayaydh2023examining,albayaydh2024innovative}, suggesting the design and implementation of a customized privacy advisor to address the unique privacy concerns and needs of MDWs~\cite{albayaydh2024co}. They have also explored how social norms, traditions, religious beliefs, and economic status influenced power dynamics within smart homes. They found that MDWs in Jordan often accepted their working conditions, including the presence of smart home devices, primarily due to their socioeconomic circumstances. Abu-Salma et al. have also shown that power imbalances were exacerbated by the use of smart home devices, with US-based MDWs frequently being both bystanders and subject of surveillance, and employers often maintained significant control over the installation, disclosure, and use of these devices, prioritizing household management and security needs over the privacy concerns of MDWs~\cite{abu2024they}. Ju et al. have examined employer-MDW relationships, particularly during the pandemic in Macau. They found that surveillance affected MDWs' sense of privacy and autonomy, influencing their experiences and perceptions of security and control in the workplace~\cite{ju2023re}. 
However, they did not investigate how power imbalances in these relationships impacted MDWs, domestic worker agencies, employers, and households within the unique cultural and social context of China. Our study is the first to explore the privacy and security needs of Chinese MDWs, with a particular focus on rural-to-urban migrants. By examining how social and cultural norms such as Confucian values and collectivism influence MDWs' privacy concerns and behaviors, our research offers a culturally nuanced perspective that has been lacking in previous studies.

\subsection{Relevant Legal Frameworks in China}
\label{subsec:regulations}
China recently enacted privacy laws and regulations aimed at protecting the privacy and personal information of its citizens, including the Cybersecurity Law (CSL)\footnote{\href{http://www.npc.gov.cn/zgrdw/npc/xinwen/2016-11/07/content_2001605.htm}{Cybersecurity Law (CSL) of the People's Republic of China}}, Data Security Law (DSL)\footnote{\href{https://www.gov.cn/xinwen/2021-06/11/content_5616919.htm}{Data Security Law (DSL) of the People's Republic of China}}, and Personal Information Protection Law (PIPL)\footnote{\href{http://en.npc.gov.cn.cdurl.cn/2021-12/29/c_694559.htm}{Personal Information Protection Law (PIPL)}}. While some provisions emphasize the privacy of personal information (e.g., Articles 22, 41-44 in CSL and Articles 1, 7, 8 in DSL), both laws primarily prioritize national security over individual privacy rights~\cite{calzada2022citizens}. PIPL specifically incorporates articles on individual privacy based on the Chinese context (e.g., Chapter 3). However, some studies have pointed out that certain provisions in PIPL—such as the definition of separate consent and the criteria for lawful anonymization and de-identification—may be difficult to enforce due to their ambiguity and the overlapping responsibilities of administrative authorities~\cite{yao2023overcoming,creemers2022china,zhang2024china}. Additionally, balancing individual privacy rights with national security needs remains a challenge, particularly when permitting government access to personal data under certain conditions (e.g., Chapter 2, Section 3)~\cite{wang2024justifying,zhou2024understanding}. We also analyzed laws governing the domestic service industry (see Table~\ref{tab:law}
), including Labor Law (LL)\footnote{\href{http://www.npc.gov.cn/zgrdw/englishnpc/Law/2007-12/12/content_1383754.htm}{Labor Law of the People's Republic of China (LL)}}, Interim Measures\footnote{\href{https://www.gov.cn/gongbao/content/2013/content_2361575.htm}{Interim Measures for the Administration of the Family Services Sector}}, Guiding Opinions on the Employee System-based Transformation and Development of the Domestic Services Industry\footnote{{\href{https://www.ndrc.gov.cn/xxgk/zcfb/tz/202312/t20231229_1362986.html}{Guiding Opinions of the National Development and Reform Commission and Other Departments on Supporting and Guiding the Employee System-based Transformation and Development of the Domestic Services Industry}}}, and local regulations such as relevant regulations in Shenzhen\footnote{\href{https://www.sz.gov.cn/zfgb/2020/gb1164/content/post_8048952.html}{Regulations of Shenzhen Special Economic Zone on Domestic Services}} and Shanghai\footnote{\href{https://sww.sh.gov.cn/dwmygl/20200706/3d5ca9e5d77f41839821315c43d8edb7.html}{Regulations of Shanghai Municipality on Domestic Services}}. 

\renewcommand{\arraystretch}{1}
\begin{table*}[!ht]
    \centering
    \fontsize{8pt}{10pt}\selectfont
    \resizebox{\textwidth}{!}{%
    \begin{tabular}{p{0.5cm} | p{3cm} p{3cm} p{3cm} p{3cm} p{3cm}}
    \toprule
        ~ & \textbf{Shenzhen Economic Zone Household Services Regulations} & \textbf{Shanghai Domestic Services Regulations} & \textbf{Guiding Opinions on Employee System-based Model} & \textbf{Interim Measures for Household Services} & \textbf{Labor Law of the People's Republic of China (LL)} \\ 
    \midrule
        \centering{\raisebox{-1\height}{\makebox[0pt][c]{\rotatebox{90}{\textbf{Regulatory scope}}}}} & In the Shenzhen Economic Zone, the policy supports the household services industry by addressing labor contracts, service agreements, and outlining the responsibilities of both service providers and consumers. & This regulation standardizes the household services industry in Shanghai, protecting workers' rights and defining the responsibilities of agencies and consumers. & This national policy seeks to shift the household service industry from a freelance model to an employee-based structure, aiming to standardize the sector and improve the welfare of domestic workers. & It establishes a framework for regulating the household services industry, including the registration of service providers, protection of consumer rights, and outlining the responsibilities of domestic workers. & It applies to all workers, including domestic workers, and establishes the fundamental legal framework for labor rights, employment contracts, wages, and working conditions. \\
    \hline
        \centering{\raisebox{-2.5\height}{\makebox[0pt][c]{\rotatebox{90}{\textbf{Strengths}}}}} & 1) It provides a comprehensive framework for service provider registration, labor contracts, and dispute resolution.
        
        2) It requires that domestic workers receive written labor contracts and prohibits employers from withholding their identification or qualification documents. & 1) Tailored to Shanghai, the regulation focuses on quality control, service standards, and the professional certification of domestic services.
        
        2) It mandates formal contracts between domestic workers and service agencies to establish clear employment terms and conditions.
        
        3) It requires consumers to respect domestic workers' dignity and ensure a safe working environment. & 1) It calls on local governments to facilitate the transition to an employee-based model by implementing policies that lower costs and mitigate risks.
        
        2) It offers financial and policy support to agencies transitioning to this model, including tax incentives and expanded social security access for domestic workers.
        
        3) It outlines long-term objectives for nationwide implementation, aiming for widespread adoption by 2035, aiming to improve job stability and professionalize the industry. & 1) It establishes a foundational regulatory framework for China's household services industry, emphasizing proper registration and oversight of service providers.
        
        2) It establishes mechanisms for dispute resolution between consumers and service providers, ensuring protections for both parties. & 1) It outlines comprehensive labor rights, including minimum wage standards, regulations on working hours, and conditions for employment termination.
        
        2) It ensures legal protections for all workers, guaranteeing fair compensation, rest periods, and safeguards against unfair dismissal.\\
    \hline
        \centering{\raisebox{-1\height}{\makebox[0pt][c]{\rotatebox{90}{\textbf{Limitations}}}}} & Although the regulation is comprehensive, its effectiveness relies on local enforcement, which can vary in quality. & The regulation is limited to Shanghai, which means its impact is localized and may not influence practices in other regions. & Transitioning to an employee system-based model may prove challenging for smaller agencies that lack the resources necessary to meet the new requirements. & The regulation is broad in scope and may lack the detailed provisions needed to address specific issues in various regions or particular areas of the industry. & The effectiveness of this law in safeguarding domestic workers depends on local enforcement, which may vary in consistency. \\
    \hline  
        \centering{\raisebox{-1.2\height}{\makebox[0pt][c]{\rotatebox{90}{\textbf{Privacy concerns}}}}} & 1) It explicitly requires service providers to avoid unnecessary intrusions into workers' privacy when monitoring their performance.
        
        2) It does not address digital privacy issues, such as how workers' personal data is managed by agencies using digital platforms. & While the regulation addresses the rights of domestic workers, it does not directly address modern privacy concerns, especially regarding the management of digital data. & While it focuses on labor rights and industry standardization, it does not specifically address how domestic workers' privacy can be safeguarded, particularly in light of the growing use of digital record-keeping and management. & There is a lack of detailed guidance on how to manage and protect the personal information privacy and security of domestic workers, especially in a digital context. & It may not fully address the unique conditions of domestic work, where workers often live in private homes and may encounter specific forms of exploitation, such as violations of their privacy.\\
    \bottomrule 
    \end{tabular}}
    \caption{Analysis of Chinese Laws and Regulations Governing the Domestic Service Industry.}
    \label{tab:law}
\end{table*}

We found that China lacks clear privacy protections for MDWs in relevant regulations. While Items 1 and 2 of Article 13 in PIPL state that personal information handlers can process personal data with individuals' consent or as outlined in employment contracts, they do not provide clarification on what constitutes surveillance in employers' homes for these workers, creating uncertainty about whether these areas are regarded as private or as workplaces. Moreover, national and local domestic service regulations---e.g., regulations in Shanghai (Article 14) and Shenzhen (Article 12)---prioritize the needs of customers and employers over the privacy of MDWs in the workplace. Although previous studies have examined legal standards regarding enforcement, public awareness, and cultural attitudes toward privacy in China, they have not fully explored how gaps in legal protections affect vulnerable groups such as MDWs~\cite{wang2024justifying,zhou2024understanding,zhang2024china}. Our paper draws attention to the gaps between legal provisions and their actual enforcement, particularly with regard to smart home technologies, and explores how these gaps pose privacy risks to MDWs. We also show how these legal gaps result in privacy ramifications for at-risk groups in China (see \textbf{\S\ref{perspectives on PSS in workplace}}, \textbf{\S\ref{mitigation practices and suggestions}}, and \textbf{\S\ref{legal suggestion}}).

\section{Methodology} \label{method}
We conducted semi-structured interviews with 26 MDWs and 5 domestic worker agencies in mainland China, using Mandarin for all interviews\footnote{Our translated interview guides, screening and exit surveys, and text copies of the codebooks (all available in English) can be found at \url{https://osf.io/nmp4b/?view_only=c4497a5eadf7485cbca0d22209067e15}.}. 
To address ethical concerns, we removed all identifiable information, including names, addresses, employer names, and specific work details, from the data set. We also assigned pseudonyms to all participants and anonymized the quotes in our reporting. Additionally, we restricted access to personal and contact details to authors directly involved in data collection and analysis. The Research Ethics Committee at King's College London reviewed and approved our study (Ethics ID: LRS/DP-22/23-35814). 

\subsection{Recruitment}
We aimed to recruit a demographically diverse sample of MDW participants who were Chinese nationals, aged 18 and older, and worked in various live-in and live-out roles (e.g., nannies, caregivers, house cleaners, maintainers) in an employer's smart home (e.g., with surveillance cameras). During recruitment, we found that the majority of MDWs (90/94) who responded to our screening survey owned smart home devices. Although we did not intend to exclude those without such devices, the final sample consisted primarily of participants who owned a smart home device in their own home. For participants working with a domestic worker agency, we recruited Chinese nationals aged 18 and older who had experience in communication, management, and negotiation with both MDWs and their employers. We recruited participants between September 2023 and March 2024 by promoting our study on online social networks, including WeChat and Baidu Tieba
, and contacting domestic worker agencies in China. We asked all interested MDWs to complete our online screening survey in Chinese. 
We also used snowball sampling to identify participants employed as staff members at domestic worker agencies in Hangzhou and Shanghai~\cite{parker2019snowball}, as these cities are key career destinations for MDWs and have a high demand for a large workforce of domestic workers~\cite{ndrc}. Given the compensation rates and time commitment\footnote{In 2021, the average monthly salary for MDWs was 6,972 yuan, marking a 21.2\% increase from 2020~\cite{mdw-salary}. With continued annual wage growth, the projected hourly wage for MDWs in 2024 is expected to be between 55 and 60 yuan, in line with current standards in urban China.}, 
we compensated participants 150 yuan ($\approx\$20.7$).

\subsection{Participants}
Table \ref{tab:MDW_demographic} presents an overview of the demographics of our MDW participants (n=26), with 13 women and 13 men. In terms of education, 6 participants had less than high school education, 5 were high school graduates, 6 had some college but no degree, 4 held an associate or technical degree, 4 had a bachelor's degree, and 1 had a master's degree. Of the 26 participants, 17 worked as live-in MDWs, while 9 worked as live-out MDWs. They held various roles, including working as a babysitter (7), house cleaner (7), nanny (6), maintenance worker (3), postnatal caregiver (3), porter (2), and elderly caregiver (1). Additionally, 21 of participants had worked with employment or domestic worker agencies. Furthermore, all participants reported having/using smart home devices in their own homes, including smart TVs (19), smart lights (5), smart speakers (12), smart doorlocks/doorbells (17), smart home cameras (19), baby cameras (1), and smart kitchen appliances (13). Further, Table \ref{tab:A_demographic} provides an overview of the demographics of our agency participants (n = 5: 3 women and 2 men). 
One participant was a high school graduate, three had some college education but no degree, and one held a bachelor's degree. Two participants had 1-3 years of work experience with agencies, one had 3-5 years, and two had over 10 years.

\renewcommand{\arraystretch}{1}
\begin{table*}[!ht]
    \centering
    \fontsize{8pt}{10pt}\selectfont
    \resizebox{\textwidth}{!}{%
    \begin{tabular}{cccccccccp{7cm}}
    \toprule 
        ~ & \textbf{Age} & \textbf{Gender} & \textbf{Education} & \textbf{Job type} & \textbf{Experience} & \textbf{Relationship w/ smart home} & \textbf{Location} & \textbf{Agency} & \textbf{Owned devices} \\ 
    \midrule
        MDW1 & 45-55 & W & Less than high school & Babysitter & 5-10 & Live-in & Zhejiang & Yes & Smart TV, smart speaker, smart camera \\
    \hline
        MDW2 & 45-55 & W & High school & Babysitter & 5-10 & Live-in & Henan & Yes & Smart TV, smart doorlock/doorbell, smart kitchen appliances \\
    \hline
        MDW3 & 45-55 & W & Less than high school & Nanny & 5-10 & Live-in & Zhejiang & Yes & Smart TV, smart kitchen appliances \\
    \hline  
        MDW4 & 25-34 & M & Bachelor’s degree & House cleaner, porter & 1-3 & Live-out & Shandong & Yes & Smart TV, smart lights, smart camera, baby camera, smart kitchen appliances \\
    \hline 
        MDW5 & 35-44 & W & High school & Babysitter & 5-10 & Live-in & Hubei & Yes & Smart TV, smart camera, smart doorlock/doorbell \\
    \hline 
        MDW6 & 18-24 & M & Some college (no degree) & House cleaner & 0-1 & Live-out & Fujian & No & Smart TV, smart doorlock/doorbell, smart camera, smart kitchen appliances \\
    \hline 
        MDW7 & 35-44 & M & Some college (no degree) & House cleaner & 5-10 & Live-out & Jiangsu & No & Smart TV, smart speaker, smart camera, smart doorlock/doorbell \\
    \hline 
        MDW8 & 35-44 & W & High school & Nanny & 3-5 & Live-in & Zhejiang & Yes & Smart TV \\
    \hline 
        MDW9 & 25-34 & M & High school & House cleaner & 3-5 & Live-out & Jiangsu & No & Smart camera, Smart doorlock/doorbell \\
    \hline 
        MDW10 & 25-34 & M & Associate/technical degree & Maintenance worker & 1-3 & Live-out & Hunan & No & Smart TV, smart speaker, smart camera, smart doorlock/doorbell \\
    \hline 
        MDW11 & 45-55 & W & Less than high school & Postnatal caregiver & 5-10 & Live-in & Gansu & Yes & Smart TV, smart speaker, smart camera \\
    \hline 
        MDW12 & 35-44 & W & Less than high school & House cleaner & 5-10 & Live-out & Zhejiang & No & Smart TV, smart lights, smart speaker, smart camera, smart doorlock/doorbell, smart kitchen appliances \\
    \hline 
        MDW13 & 25-34 & M & Some college (no degree) & House cleaner & 1-3 & Live-out & Jiangsu & No & Smart TV, smart lights, smart speaker, smart doorlock/doorbell, smart kitchen appliances \\ 
    \hline 
        MDW14 & 25-34 & M & Associate/technical degree & Maintenance worker & 3-5 & Live-out & Jiangsu & No & Smart speaker \\ 
    \hline
        MDW15 & 18-24 & M & High school & Elderly caregiver & 0-1 & Live-in & Shandong & Yes & Smart speaker, smart camera, smart kitchen appliances \\
    \hline 
        MDW16 & 35-44 & M & Bachelor’s degree & Porter & 3-5 & Live-out & Fujian & No & Smart speaker, smart doorlock/doorbell \\
    \hline 
        MDW17 & 35-44 & W & Less than high school & Babysitter & 1-3 & Live-in & Fujian & Yes & Smart camera, smart doorlock/doorbell \\
    \hline 
        MDW18 & 25-34 & M & Associate/technical degree & House cleaner & 3-5 & Live-out & Chongqing & Yes & Smart TV, smart camera, smart doorlock/doorbell, smart kitchen appliances \\
    \hline 
        MDW19 & 25-34 & M & Some college (no degree) & Nanny & 3-5 & Live-in & Fujian & Yes & Smart TV, smart speaker, smart camera, smart doorlock/doorbell, smart kitchen appliances \\ 
    \hline 
        MDW20 & 25-34 & M & Some college (no degree) & Nanny & 1-3 & Live-in & Hebei & No & Smart TV, smart speaker, smart camera \\
    \hline 
        MDW21 & 25-34 & W & Bachelor’s degree & Nanny, babysitter & 1-3 & Live-in & Shanghai & Yes & Smart camera \\
    \hline 
        MDW22 & 45-55 & W & Associate/technical degree & Postnatal caregiver & 10+ & Live-in & U.S.A & Yes & Smart TV, smart lights, smart speaker, smart camera, smart doorlock/doorbell, smart kitchen appliances \\
    \hline 
        MDW23 & 35-44 & W & Master’s degree & Nanny, babysitter & 1-3 & Live-in & Shanghai & Yes & Smart camera, smart doorlock/doorbell, smart kitchen appliances \\
    \hline 
        MDW24 & 25-34 & M & Bachelor’s degree & Maintenance worker & 5-10 & Live-out & Hebei & Yes & Smart TV, smart camera, smart doorlock/doorbell, smart kitchen appliances \\
    \hline 
        MDW25 & 45-55 & W & Less than high school & Postnatal caregiver & 10+ & Live-in & Shanghai & Yes & Smart TV, smart lights, smart doorlock/doorbell \\
    \hline 
        MDW26 & 35-44 & W & Some college (no degree) & Babysitter & 5-10 & Live-in & Zhejiang & Yes & Smart TV, smart speaker, smart camera, smart doorlock/doorbell, smart kitchen appliances \\
    \bottomrule 
    \end{tabular}}
    \caption{MDW participant demographics, work details, experience working with domestic worker agencies, and owned smart home devices. `Location' indicates the work and residence area, and `Agency' refers to employment through domestic worker agencies.}
    \label{tab:MDW_demographic}
\end{table*}

\renewcommand{\arraystretch}{1}
\begin{table*}[!ht]
    \centering
    \fontsize{8pt}{10pt}\selectfont
    \resizebox{\textwidth}{!}{%
    \begin{tabular}{cccccccccc}
    \toprule
        ~ & \textbf{Age} & \textbf{Gender} & \textbf{Education} & \textbf{Experience} & \textbf{Location} & \textbf{Full-time/part-time} & \textbf{Clients served} & \textbf{Agency size} & \textbf{MDWs employed} \\  
    \midrule
        A1 & 25-34 & M & Associate/technical degree & 1-3 & Zhejiang & Full-time & 720+ & 101-500 & 101-500 \\
    \hline
        A2 & 35-44 & W & Some college (no degree) & 5-10 & Zhejiang & Full-time & 1000+ & 1000+ & 1000+ \\ 
    \hline
        A3 & 35-44 & M & Some college (no degree) & 5-10 & Zhejiang & Full-time & 1000+ & 1000+ & 1000+ \\ 
    \hline  
        A4 & 35-44 & W & High school & 3-5 & Gansu & Full-time & 1500+ &  501-1000 &  1000+ \\ 
    \hline 
        A5 & 25-34 & M & Bachelor’s degree & 1-3 & Shanghai & Full-time & 100+  & 11-50 & 51-100 \\ 
    \bottomrule
    \end{tabular}
    }
    \caption{Demographics and background information of staff members of domestic worker agencies.}
    \label{tab:A_demographic}
\end{table*}

\subsection{Interview Procedure}
Four authors, all native Mandarin speakers, conducted the interviews remotely in Mandarin using VooV Meeting\footnote{We chose \href{https://voovmeeting.com/}{VooV Meeting}, a widely used Tencent video conferencing tool in China, for its accessibility, participant familiarity, minimal setup, and encrypted communication to ensure confidentiality and anonymity.}, with an average interview duration of 83.4 minutes for MDWs and 63.6 minutes for agencies. We reached data saturation after 22 interviews, and then conducted four additional interviews to confirm that saturation had been reached. Four authors continuously reviewed the data after each interview, observing that insights started to repeat, and that additional interviews did not introduce new insights~\cite{guest2006many}. We audio-recorded and transcribed all interview sessions using notta.ai~\cite{nottaai2024}. 
All interviewers carefully reviewed and refined the transcripts to ensure accuracy and reduce cross-language inconsistencies~\cite{squires2009methodological}.

We started the interviews by asking participants to describe their work duties and professional experiences as MDWs. We explored their privacy and security concerns while working in employers' smart homes, along with their perspectives on power imbalances in their relationships with both agencies and employers. Additionally, we investigated their experiences with smart home devices, privacy concerns, feelings of intrusion, and interactions within their own smart homes (e.g., parental monitoring vs. a child's privacy needs). Finally, we asked participants to share their definitions and expectations regarding privacy and security, along with recommendations to improve existing legal guidance from an MDW perspective. 

Interviews with domestic worker agency participants began with discussions about their backgrounds and responsibilities (e.g., mediating negotiations between MDWs and clients) and their perspectives on existing contracts and training, particularly regarding privacy and security practices. We then explored their views on power imbalances in client-MDW relationships, examining how they supported MDWs and/or clients in privacy and security disputes (e.g., handling clients' private data storage). Additionally, we gathered their opinions on privacy and security laws related to MDWs and their labor and privacy rights. Finally, all agency participants completed an exit survey after the interviews to provide their demographics.

\subsection{Pilot Study}
We conducted four pilot tests (three with MDWs and one with an agency) to collect feedback and suggestions for improvement. These pilot interviews were not included in the final data analysis. The feedback we received helped us refine several interview questions, including updating those related to differences in attitudes between live-in and live-out MDWs (e.g., we revised questions about live-in situations that might not be relevant to live-out MDWs). We also revised questions concerning MDWs' legal knowledge and expectations, their views on power imbalances (e.g., perspectives on contract issues), and how their attitudes evolved over time (e.g., whether their privacy concerns changed, and how this impacted their practices at home). Additionally, we added questions about private data protection practices within agencies (e.g., how agencies collected and stored both MDWs' and clients' data). Further, we updated questions on power imbalances in client-MDW relationships and added new questions to gain a deeper understanding of agencies' views on the privacy of both MDWs and their clients.

\subsection{Data Analysis}
To analyze our qualitative interview data, we employed an inductive thematic analysis approach~\cite{braun2006using}. For MDW interview transcripts, four authors independently coded the same randomly selected MDW interview transcript and created separate codebooks. They then met to resolve disagreements and merge their codebooks into one. Using this merged codebook, the same four authors independently coded two more MDW interview transcripts to evaluate its effectiveness, followed by discussions and iterations. This process was repeated one more time until code saturation was achieved. 
Using the finalized codebook, the first author coded all 26 MDW interview transcripts, while two other authors each coded ten transcripts, and another coded six transcripts. Inter-rater reliability was measured, resulting in an average Cohen's kappa of 0.88 for author 1 and author 2, 0.82 for author 1 and author 3, and 0.81 for author 1 and author 4, indicating excellent agreement~\cite{fleiss2013statistical}. The authors then organized the codes into themes to address our main research questions. For agency interview transcripts, authors 1 and 2 coded the same transcript, developed separate codebooks, resolved conflicts, and merged the codebooks into one. Using this merged codebook, they coded the remaining four transcripts to achieve code saturation, discussed any disagreements, and finalized the codebook. The authors then coded the five agency interview transcripts using the finalized codebook, resulting in an average Cohen's kappa of 0.86, indicating excellent agreement~\cite{fleiss2013statistical}. They then organized the codes into themes to address our main research questions. 

\subsection{Author Positionality} \label{appendix:positionality}
Privacy, as a psychological concept, is defined as the way people regulate personal boundaries and implement strategies to control access to their personal space and information~\cite{westin1968privacy,altman1975environment,petronio2002boundaries}. It can vary significantly based on cultural (e.g.,~\cite{kim2023privacy,benamati2021information}), social (e.g., individualism vs. collectivism~\cite{johnston2009national,li2017cross}), and individual (e.g., self-disclosure and personality traits~\cite{skirpan2018s,meier2024privacy}) factors. Instead of providing participants with a predefined concept of privacy, largely shaped by WEIRD values (e.g., individualism), we chose to examine participants' privacy and security perspectives and practices through the lens of their individual backgrounds, experiences, and perceptions, while also considering distinct cultural and social factors in China, such as collectivism and Confucianism. Although participants did not explicitly mention \textit{``Confucianism''} during the interviews, they frequently referred to concepts and values related to it, such as family harmony and filial piety (see \textbf{\S\ref{chinese_cultural_and_social_factors}}). 
Confucianism emerged organically through participant discussions and the cultural context reflected in their responses. Instead of being a specific code in our codebook, Confucianism emerged as a broader theme during the data analysis process.

In addition, our interpretation of the results is influenced by our background and experience. All authors have extensive expertise in human-centered computing and have conducted research with both majority populations and at-risk groups. Although we are trained researchers based in predominantly Western institutions, four authors, including interviewers and data analysts, are native Mandarin-speaking Chinese and have lived or currently live in China. This shared cultural background allowed us to identify and interpret culturally specific nuances in the data that could not have been obvious. One author initially translated our codes and quotes, and the other three authors reviewed them to ensure the translation's accuracy and authenticity.

\subsection{Limitations}
One limitation of our study is the exclusive use of qualitative methods. Although we initially considered a quantitative approach to capturing broader trends and statistically significant data on domestic workers' privacy concerns and practices in multi-user smart homes, we ultimately determined that qualitative methods were better suited to uncovering the nuanced, context-specific insights that quantitative methods might overlook. Our primary goal was to explore the complex, context-dependent experiences of MDWs in multi-user smart homes. Using qualitative methods, we aimed to generate new insights into the unique challenges and perspectives of MDWs that would not be easily captured through statistical analysis (see \textbf{\S\ref{sec:discussion}}). Although recruitment challenges played a role in our decision, the qualitative approach enabled us to conduct an in-depth exploration of participants' lived experiences and perspectives~\cite{soden2024evaluating}. 
Additionally, our research relies on self-reported data, which is shaped by participants' perceptions, memories, and interpretations, potentially introducing biases such as social desirability bias~\cite{nederhof1985methods}, where participants might adjust their responses to appear more favorable. Future research should incorporate a more diverse sample and utilize quantitative methods to explore privacy issues in different settings.
\section{Results} \label{results}
In this section, we present our findings. 
\S\ref{awareness and perspectives} presents the views, attitudes, and practices of our MDW participants (RQ1). \S\ref{perspectives on PSS in workplace} describes privacy concerns, power dynamics, and challenges in the workplace, reflecting how these experiences affected relationships with agencies and employers (RQ2). \S\ref{mitigation practices and suggestions} presents mitigation practices to address privacy-related conflicts at home and work, and legal perceptions and suggestions to improve privacy and security (RQ3). Although MDW participants frequently shared first-hand accounts of their lived experiences with regard to privacy and security in households, our agency participants offered insights into broader structural and procedural factors—such as recruitment, training, and conflict resolution—that influenced these experiences. Participants' views were interconnected, highlighting the relational dynamics between the two groups. By presenting these views together, we aim to provide a holistic understanding of how privacy and security concerns are shaped, discussed, and addressed within the context of domestic work.

\subsection{Individual Understanding of and Perspectives on Privacy and Security} \label{awareness and perspectives}

\subsubsection{Definitions of and attitudes toward privacy and security} \label{definition of PSS}
We present MDW participants' interpretations of privacy and security, their attitudes toward these concepts, and their approaches to seeking related advice.

\textbf{Definitions of privacy and security.} 
Participants defined \textbf{privacy} through both physical and digital dimensions, emphasizing the protection of personal information, as illustrated by MDW15, \textit{``In addition to ID, my personal privacy also includes bank account, social insurance number, Alipay or WeChat password, and my face and fingerprints.''} Interestingly, some participants were not concerned about the collection of certain personal information, such as phone numbers, ID numbers, and facial data. This distinction highlights how participants prioritized privacy concerns based on perceived risk, potentially shaping their responses to data practices and security measures in smart home environments. As MDW18 stated: 
\textit{``I think these minor privacy infringements in daily life do not significantly impact me, so I don't think it's a big deal if these devices collect my ID and phone numbers [...] Privacy becomes a concern only when it involves personal safety or property; then it can be deemed unsafe and intrusive.''} 

More than half of our participants associated personal information \textbf{security} with property protection, physical safety, and privacy. They perceived privacy and security as an interchangeable concept~\cite{colnago2023there}, describing security as protecting personal data (e.g., preventing financial fraud), personal physical safety, and property safety. MDW10 stated: \textit{``Privacy refers to personal information that one does not want to disclose [...]. Security means ensuring that this information is not leaked or misused, ensuring that individuals are not infringed upon or harmed.''}


\textbf{Attitudes toward excessive data collection and normalization of data breaches.} About one-third of participants pointed out the extensive collection of personal data by smart devices and apps, often exceeding what was required, which raised major privacy concerns. 
For instance, MDW23 discussed the balance between privacy needs and UX: \textit{``
If you don’t give these devices and apps permissions, they won’t work, forcing you to authorize it [...]. This feels intrusive, but to enjoy the convenience, you have to sacrifice some privacy.''} Additionally, some participants discussed how normalization of data breaches could lead to survivorship bias~\cite{brown1992survivorship}, where normalization of data collection and perceived benefits of smart devices could overshadow privacy concerns, leading users to undervalue the importance of data protection until they personally experience a breach. As MDW15 stated: \textit{``Everyone is using these devices, and there is definitely a risk of data leakage with these devices. However, I feel quite safe, which might just mean that my data hasn't been leaked yet.''}

\textbf{Strategies for seeking and sharing privacy-related advice.} Almost one-third of participants actively sought privacy-related advice by independently researching online sources, such as news feeds and social media, or consulting acquaintances. 
For example, MDW6 used consultation hot-lines like 12345\footnote{The Chinese dealt with all of these problems by setting up the \href{https://www.economist.com/the-economist-explains/2017/02/07/what-are-chinas-12345-hotlines}{12345 helpline} in 1987. The helpline is a phone and online service available to anyone in China, allowing them to ask questions or file complaints.} to seek advice, while MDW17 preferred to use government information sources: \textit{``I follow a few government and public security accounts that have shared tips on managing permissions for smart speaker apps and potential eavesdropping. These accounts are fairly reliable, unlike personal marketing accounts or those just trying to get popular. Official information is definitely more reliable.''} Furthermore, several participants highlighted the importance of the National Anti-Fraud Center (NAFC) app\footnote{\href{https://www.gjfzpt.cn}{The National Anti-Fraud Center (NAFC)} is a Chinese mobile app developed by the Ministry of Public Security. Its purpose is to safeguard the telecommunication network, provide channels for reporting online fraud, and enhance citizens' awareness of fraud prevention.} to address financial security concerns. 
However, MDW10 noted the lack of NAFC initiatives focused on the security of smart home devices.


\subsubsection{Social views on privacy and security} \label{social view} We present MDW participants' perspectives on social factors (e.g., public surveillance, verification of real names) and how demographic differences (e.g., gender, age) influenced their perceptions of privacy.

\textbf{Views on public monitoring and household surveillance.} Most participants regarded public surveillance (e.g., CCTV cameras) as necessary for safety and order, with many showing indifference or acceptance. They mentioned that their trust in the government and public authorities made them less concerned about privacy. As MDW9 stated: \textit{``CCTV cameras are everywhere now, and people don’t really notice them anymore [...] Since they’re mostly owned by the government or public security, they’re seen as a way to protect property and provide evidence if something happens, so privacy worries are pretty low.''}

In contrast, many participants found private monitoring in employers' homes invasive, expressing discomfort. MDW22 pointed out the differences between China and the United States, noting \textit{ ``In China, cameras are common in residential and public areas for safety, while in the US, the focus is on protecting individual property rather than public.''} Furthermore, some participants noted that surveillance in public workplaces, such as daycare centers, felt less invasive compared to constant monitoring in employers' homes. For example, MDW11 mentioned feeling increased pressure and self-consciousness when subjected to surveillance in employers' homes: \textit{``At home, it feels like the camera is just monitoring me, but in public, the camera is watching everyone, so it feels less pressured.''} 

\textbf{Private data leakage due to real-name verification.} 
Participants discussed the risks of data leakage from smartphones and apps, linking widespread fraud and nuisance calls to the exposure of phone numbers due to verification requirements of real names in certain apps. 
MDW22, a live-in postnatal caregiver who worked in both China and the US, stated: \textit{``In China, phone numbers are linked to IDs due to real-name requirements, unlike in the US where phone cards are not tied to real names and apps, like WhatsApp, generally require an email address rather than a real-name phone number, reducing my privacy breach risks.''} Similarly, MDW7 pointed out that real-name registration on digital platforms increased the accessibility of personal data to hackers. MDW3 further noted, \textit{``While ID numbers and facial recognition data are secure due to real-name verification, phone number breaches remain a severe issue for Chinese citizens.''}

\textbf{Gender effects.}
We found that privacy perceptions varied among participants based on gender, with women generally expressing greater concern about privacy issues than men. For example, they were more proactive in protecting their privacy and more often faced privacy violations, harassment, or judgment, as noted by MDW24: \textit{``Live-in women MDWs are especially concerned about their physical privacy being compromised in employers' homes. But once privacy leaks, sometimes they are ashamed to tell.''} In contrast, MDW4 noted, \textit{``Men, especially live-out MDWs, often perceive privacy breaches as less significant, potentially due to physical biometric advantages that offer them a greater sense of security.''} 

\textbf{Age and personality influenced privacy awareness.}
Participants observed that factors related to age and personality traits significantly affected privacy concerns. Younger MDW participants were more tech-savvy and therefore more cautious and privacy conscious, while older participants tended to have less awareness of the technical aspects of privacy breaches. As MDW22 shared, \textit{``Younger people, being more tech-savvy, are more cautious and aware of privacy issues due to awareness of digital privacy invasions.''} Furthermore, participants with relaxed and easy-going personalities often exhibited lower privacy concerns, possibly due to their tendency to avoid evaluating the pros and cons of self-disclosure and to refrain from over-analyzing potential risks~\cite{ostendorf2022theoretical,meier2024privacy}. As MDW7 stated, \textit{``I think people with meticulous or detail-oriented personalities are more likely to be aware of and take steps to protect their privacy.''}

\subsection{Privacy Concerns, Challenges, and Power Dynamics in the Workplace} \label{perspectives on PSS in workplace}
\subsubsection{Privacy concerns and needs in employers' homes} \label{concerns in workplace}
We present MDW participants' perceptions and awareness of workplace monitoring, including its impact on their performance. Additionally, we highlight the differences in surveillance experiences and perspectives between live-in and live-out MDWs.

\textbf{Privacy concerns and discomfort in the workplace.} Half of MDW participants perceived camera-based smart home devices, like surveillance and baby monitors, as significant privacy issues. They shared that the constant feeling of being watched caused anxiety, distrust, and frustration, especially when cameras were placed in private areas such as bedrooms. 
For example, MDW15, a live-in elderly caregiver, described the discomfort and unease that mobile surveillance devices caused in their daily life: 
\textit{``Sometimes I feel like this monitoring robot is always following me. I know my employer uses it to monitor me, which makes me uneasy. While working, it sometimes stays beside me and speaks unexpectedly, which can be scary [...] When he checks my phone's location, I feel constantly monitored, even when I'm out with friends. This invades my privacy and makes me feel very restricted.''}


In contrast, all agency participants acknowledged the surveillance concerns raised by MDW participants and made efforts to balance the security needs of employers with the privacy rights of workers. A1 and A4 expressed empathy and emphasized the importance of transparent communication regarding the placement and purpose of cameras. They highlighted the need to establish clear boundaries between employers and MDWs to safeguard workers' privacy. 
As A4 described: \textit{``We will communicate with clients to ensure that the cameras in their homes comply with our regulations. Our company does not allow clients to install cameras in MDWs' bedrooms.''}

\textbf{Perceptions and awareness of monitoring in employers' homes.} Most MDW participants noted that although surveillance devices were framed as safety measures, they were primarily seen as tools for monitoring. This created an understanding among MDW participants that they were constantly being watched. They emphasized a strong preference for clear communication and transparency about the use of these devices. The absence of explicit communication led to feelings of distrust, as MDW participants believed that the true purpose of the devices was not fully disclosed~\cite{kim2023privacy,abu2024they}. 
Some participants, like MDW22, argued that being aware of constant monitoring could undermine trust and affect workplace interactions, causing feelings of discomfort and self-consciousness~\cite{foucault2020panopticism}. 

In addition, most MDW participants discussed how monitoring positively affected their performance. 
For instance, MDW18 stated, \textit{``I don't worry about cameras; they're common now. We maintain high standards. Trying to hide raises suspicion. Why hide if my work is honest and straightforward?''} 
In contrast, some participants noted that monitoring created unease and negatively affected their performance. They became more cautious and stressed, striving to avoid mistakes. MDW6 noted, \textit{``Every action and word is scrutinized, making it difficult to relax or make mistakes. Without cameras, we can work at our own pace and take breaks as needed.''} 

\textbf{Impacts of monitoring on live-in and live-out MDWs.} We found that live-out MDW participants, particularly those in temporary roles, felt monitored mainly during working hours, which contributed to feelings of discomfort and pressure. 
In contrast, live-in MDW participants reported a greater impact on their privacy and autonomy due to continuous household monitoring, particularly in intimate spaces such as bedrooms. Although some recognized the need for safety—especially when caring for infants—the absence of clear communication from employers about the purpose of monitoring sparked distrust and discomfort. MDW5, a live-in babysitter, noted: \textit{``It's understandable if the camera is aimed at the baby, but it's excessive if aimed at my bed. I feel monitored and uncomfortable sleeping. I may unknowingly change clothes under the camera and still feel watched even with the door closed, especially knowing that male family members can view the recordings. This is embarrassing and violates my privacy.''} Furthermore, both groups recognized the benefits of surveillance cameras, viewing them as \textit{``proof''} that could showcase their professionalism and behavior and at times act as evidence in the event of employment disputes.

\subsubsection{Challenges and concerns about agencies} \label{agency concerns}
We present the perspectives and recommendations of MDW participants regarding contract transparency (e.g., detailed monitoring practices), their training experiences (both professional and privacy-related), the role of agencies in mediation and negotiation, and their preference for traditional agencies over digital platforms.

\textbf{Contracts and transparency.} 
Many MDW participants shared their experiences with the tripartite contracts involving agencies and employers. 
They emphasized the necessity of transparency
, like MDW17 stated: \textit{``I believe laws, or at least contracts, should specify and balance our work and life clearly, which is a fundamental labor right, 
with clearly defined rights and boundaries.''} However, some participants pointed out the lack of insurance and safety provisions in their contracts. For example, MDW15 stated, \textit{``Most workers are only covered by basic accident insurance, which does not fully address all potential risks associated with our work.''} 
Although acknowledging the importance of contracts in defining employment terms, some participants pointed out that complex legal language often made some terms difficult to understand, particularly for MDWs with limited education or legal knowledge. As MDW20 stated: \textit{``The length of contracts is overwhelming, and their complexity makes it hard to focus on the content. This can cause us to miss important details that affect our rights.''}

While recognizing the employment challenges MDWs faced, A4 noted that agencies had limited resources to protect their data. This was due to the fact that most MDWs had contracts with at least one agency, making their personal data accessible to all domestic worker agencies. A5 also raised concerns about MDWs' personal data being available through official channels: \textit{``Our company is linked to the national public security system, so we check the identity and background of MDWs, including their police records. Before 2018, you could even pay 100 yuan to get anyone's home address, criminal history, and hotel check-in records from the police [...]. Now, we do rolling checks every 15 days to screen out problematic MDWs. Those individuals are blacklisted and not allowed to work.''}

\textbf{Lack of privacy and legal training programs.} 
Half of our MDW participants highlighted that their training often focused on professional skills over privacy protection. For instance, MDW17 mentioned, \textit{``Our training covers practical skills like nursing and childcare, but no training about addressing privacy concerns.''} 
Some participants further emphasized the importance of integrating legal training, particularly for less-educated MDWs. As MDW11 stated: \textit{``If these contents can be popularized in training, even illiterate people can have a basic awareness of protection. At least, they will know how to protect themselves and how to handle privacy breaches.''}

We found that most agencies provided limited training for privacy issues related to surveillance. While some agencies considered incorporating privacy protection into their training courses, cost constraints prevented it from becoming a regular component of their programs, with a stronger focus placed on job performance and conflict resolution instead. A5 highlighted the unnecessary perceived benefit of such privacy training, viewing it as complicating the employment relationship: \textit{``If we provide privacy-related training to these domestic workers, they will become more aware of their privacy and may start asking employers for higher wages or better privacy protection. This would increase our workload and create many new issues.''} In contrast, A4 noted their agency provided legal and privacy-related training by inviting legal experts, such as lawyers from labor unions: \textit{``They (legal training) address security and contract-related issues, such as self-protection, labor rights, and procedures for addressing conflicts or injuries on the job [...] Although we offer these courses and lectures, due to the relatively low educational level of these domestic workers, they do not listen very attentively and generally go through the motions. As a result, these training courses are unlikely to produce positive feedback.''}

\textbf{Role of agencies in mediation and negotiations.} Some MDW participants emphasized the supportive role of agencies in mediating between them and their employers, particularly in fostering clearer communication and transparency regarding the use of surveillance devices. As MDW18 stressed, \textit{``Agencies should help us negotiate compensation and apologies from clients if our privacy is violated. It’s important to have someone to mediate and help maintain our dignity and rights.''} However, nearly half of MDW participants expressed frustration over agencies who prioritized client needs at the expense of the well-being and rights of MDWs. 
For example, MDW5 noted: \textit{``No agency has ever provided privacy-related content or mentioned it in contracts. Agencies focus on protecting their interests, not our privacy.''}

\subsubsection{Power imbalances in MDW-employer relationships} \label{power dynamics}
MDW participants often faced unclear work requirements and constant monitoring without consent, exacerbating power imbalances that negatively impacted their emotional and psychological well-being. MDW18 highlighted \textit{``fear of retaliation or job loss.''} MDW21 compared their experience with historical racial injustices: \textit{``The first three months were mentally tough; I felt a strong sense of inferiority. My wealthy employer treated me as a subordinate, not an equal. I felt like a black American during segregation, constantly scrutinized and undervalued by others.''}

Some live-in MDW participants mentioned that employers maintained significant control over work environments and assignments, making unilateral decisions without their input. As MDW1 stated: \textit{``The employer assigned me many tasks on the first day and asked me to complete all within four hours. I wasn't familiar with the place, so I made some mistakes. When she returned, she scolded me harshly.''} 
Participants noted their limited power to negotiate or advocate for their rights. For example, MDW26 described how this imbalance extended to employment status: \textit{``One client didn't want to pay me. After two or three months of work, they were looking for excuses to make me leave. They falsely accused me of theft, planted valuable jewelry in my suitcase, and called the police to avoid paying my salary. Our group is very passive; if falsely accused, we have no power to refute. Even if you ask for evidence, they usually won't provide it.''}

Despite legal protections, discriminatory practices persisted, complicating the ability of MDWs to seek compensation or legal recourse. For example, MDW25 discussed discrimination based on health conditions: \textit{``National laws prohibit employers from discriminating against employees with hepatitis B. They can't refuse to hire you because of this condition, but clients and agencies often reject us by questioning our health checks, claiming it affects the baby. This creates an unequal relationship.''} MDW21 highlighted how some employers and agencies used third parties to shift responsibility, creating additional barriers for MDWs seeking justice: \textit{``If a workplace accident occurs, the employer isn't liable; the third-party agency is responsible for any work injury or labor dispute. These agencies specialize in such matters and can easily disappear by canceling a business license, so when they shut down, an MDW has no chance to sue.''}

\textbf{Agency perspectives on power dynamics and rights.} Most of agency participants acknowledged the power imbalances present in employer-MDW relationships. For example, A1 pointed out that the lack of clear legal frameworks for contract definitions further reinforced the passive position of MDWs: \textit{``One client's baby hurt their nose, so they blamed the nanny. Although the video showed that she did nothing wrong, the client can terminate the contract whenever they want. No law regulates this behavior, and the client can directly inform the worker to leave their home.''} A2 even stated that this lack of clarity caused MDWs to \textit{``have no basic human rights''}. In addition, agency participants highlighted that employers often dictated employment terms and daily life, leaving MDWs with little autonomy. For example, A3 pointed out: \textit{``Employers provide MDWs with a real-time GPS tracker, making them feel very restricted, like being in prison. I believe that technological advancements might make this employment relationship more unequal and easier to monitor.''} 
However, A5 did not see these power imbalances as problematic, stating, \textit{``I don't believe there is any inequality in the relationship; it's purely an employment relationship. The client has needs, and the domestic worker provides services, exchanging labor for money.''}

\subsection{Privacy Mitigation Practices and Suggestions} \label{mitigation practices and suggestions}
\subsubsection{Privacy concerns and mitigation strategies in own home} \label{concerns in own home}

Most MDW participants prioritized freedom and autonomy in their own homes, placing usability and convenience above privacy concerns~\cite{wong2023broadening}. Their experiences with smart home devices reflected patterns similar to those in Western countries, such as valuing convenience more than privacy and security~\cite{li2021motivations,zheng2018user}. Participants generally trusted established brands over smaller ones due to their perceived reliability, safety features, and good reputation~\cite{liu2022privacy}. However, some participants mentioned that their exposure to surveillance devices in employers' homes led them to be more mindful of privacy when considering the installation of similar devices in their own homes. MDW26 stated, \textit{``In the employer's home, I am monitored, but at home, I monitor others, playing the role of the employer. I consider my children's feelings and generally avoid checking the camera unless there's a special situation.''} 

Most MDW participants emphasized the importance of managing devices and controlling permissions for privacy protection, pointing out the advantages of centralized control, typically handled by male family members with full administrative access. In these gender-based hierarchies, men in the household usually take charge, while other members, particularly women, are expected to comply without having influence on decisions~\cite{rosenlee2023gender}. For instance, MDW16 described: \textit{``I’m the one who handles the door lock settings. My wife can see them, but she doesn’t need to make changes, and only I know the admin password. When we bought the lock, it was my decision—she thought I was wasting money at all.''}

\subsubsection{Privacy mitigation strategies in employers' homes} \label{mitigation in workplace}
We describe MDW participants' workplace privacy practices and the strategies they used to negotiate privacy concerns with their employers. Additionally, we discuss how agencies prioritized the protection of clients' private information, sometimes at the expense of MDWs' privacy.

\textbf{Strategies for protecting against monitoring.} Most of MDW participants used various strategies to protect their privacy from pervasive monitoring in their employers' homes. For example, MDW12 said, \textit{``I made important calls from the balcony or outside to avoid being overheard by my employers. Sometimes I used my dialect during personal calls to prevent employers from understanding.''} More than half of participants accepted monitoring as an inevitable part of their job and sought ways to co-exist with it. 
However, a few participants acknowledged the limitations of these protective measures, as MDW17 argued, \textit{``Constantly avoiding cameras isn't feasible and might even look suspicious. It's like the saying, `No 300 taels of silver are buried here\footnote{\href{https://www.ewccenter.com/no-300-taels-of-silver-are-buried-here/}{No 300 Taels of Silver Are Buried Here (此地无银三百两)}}', you are trying too hard to hide something that can actually draw attention and expose you.''} Additionally, a few participants mentioned that some employers were considerate enough to adjust monitoring practices to respect MDWs' privacy. For example, MDW1’s employer turned off the camera in the bedroom when the baby was not present. MDW11 also shared a positive experience: \textit{``A few of my previous employers were very considerate of my privacy. During the interview, they took the time to inform me about the locations of the cameras at their home. This gesture showed their respect for my privacy, and I really appreciated it.''}

\textbf{Negotiating workplace privacy: direct communication vs. resigned acceptance.} We found that the contrast between proactive negotiation and resigned acceptance highlighted the varying degrees of agency that MDW participants had when addressing privacy concerns in the workplace. Despite this, most participants emphasized the importance of directly addressing privacy concerns related to workplace monitoring through open communication with their employers. As MDW12 highlighted: \textit{“We often address discomfort in monitoring by communicating directly with employers, or through an agency, which would then relay the concerns to the employer.”} 
However, some participants described a more resigned approach, opting to avoid conflict and adhere to employer expectations despite their discomfort. As MDW10 described: \textit{“While we understand the employer's monitoring requirements, we have no choice but to endure it for the sake of our livelihood [...]. If you want to negotiate with them, this could negatively affect the employer's opinion, leading to poor reviews or resentment [...]. If you ask them to turn off cameras, they might feel you don't trust them. Therefore, even if it bothers me, I might choose not to ask to avoid conflict.”}

\textbf{Agencies prioritized clients' data protection over MDWs’ data.}
We found that all agency participants prioritized protecting clients' data over that of MDWs, focusing on cloud storage, encryption, and restricted data access. A1 and A3 relied on cloud storage and encryption to protect personal data, with the aim of reducing risks of unauthorized access. A3 emphasized, \textit{``The potential loss of trust and legal consequences of clients' data breaches are significant concerns for us.''} A2 highlighted their comprehensive approach to client data protection, stating: \textit{``Client data is handled with the utmost care, with multiple layers of security to prevent data breaches, including firewalls, intrusion detection systems, and regular IT security audits.''} In addition, some agencies implemented strict data access policies to ensure security, as A1 noted, \textit{``Only authorized personnel have access to clients' sensitive data and contract information. We regularly review access permissions to maintain client data security.''} A5 further explained that client data protection involved both technical measures and organizational policies, which included regular employee training for the importance of data protection and compliance with regulations. Interestingly, some agencies, like A2 and A3, deployed cameras in clients' homes to monitor MDWs, aiming to ensure safety and transparency. However, even with the consent of clients, this constant monitoring caused discomfort for both clients and MDWs. A3 raised significant concerns about data leakage, noting: \textit{``Small agencies with limited budgets may fail to protect recorded footage adequately, making it vulnerable to hackers. Also, cameras might capture intimate or sensitive moments, raising serious privacy concerns.''}.

\subsubsection{Impact of collectivism and Confucian values on privacy perceptions.} \label{chinese_cultural_and_social_factors}
We found that collectivist social norms and distinct Confucian values shaped the privacy perceptions of MDW participants and their families. 
In a collectivist society like China, where communal living and shared responsibilities are the norm, there is a stronger emphasis on the needs of the community, often at the expense of individual privacy~\cite{hofstede1984culture,hofstede2011dimensionalizing}. 
Furthermore, some participants noted that privacy breaches had more significant consequences for individuals with high public profiles or substantial assets, reflecting collectivist social hierarchies. High-status groups may prioritize individualistic values, amplifying the impact of privacy breaches, while lower-status groups are more likely to embrace collectivist values~\cite{iacoviello2019collectivism}. As MDW17 noted: \textit{``Rich people, celebrities, and officials have a lot more at stake with privacy. A breach could mean financial loss, damaged reputation, or even personal safety risks, making any invasion a bigger deal for them. They need to protect their image and security, so breaches hit them hard. But for people like us, regular folks without fame or wealth, privacy still matters, but it doesn't feel as severe if it's breached.''} 

We found that Confucian values, such as family harmony and filial piety, played a significant role in shaping privacy perceptions\footnote{Although neither our MDW nor agency participants explicitly mentioned Confucianism during the interviews, it emerged as a cultural theme in our analysis. Additionally, we took into account how authors' cultural backgrounds could have influenced the identification and interpretation of themes (see `Author Positionality' in \S\ref{method}).}. 
Many MDW participants were open to using monitoring devices for caregiving, recognizing the security and safety benefits of smart home cameras for elderly parents. However, some elderly family members initially opposed the idea of being monitored, feeling uneasy about the concept of surveillance. For example, MDW4 explained: \textit{``Cameras can help monitor the house but might make the elderly feel like they're being watched. So we definitely first ask them for their opinion. If they agree, we install; if not, we don't.''}. 
These elderly family members often accepted the devices once they understood the benefits, reflecting a change influenced by filial piety, the cultural expectation that children care for their parents~\cite{wei2013confucian}. Some MDW participants felt it was their duty to ensure the safety of their parents without seeking explicit permission. As MDW9 stated: \textit{``As children, installing these devices is also a way to express our filial piety, respect, and love for our parents.''} MDW17 further acknowledged the need for \textit{family harmony}: \textit{``I just hope for family harmony [...] so I don't see privacy violations as such a big deal in our daily lives.''} Additionally, younger family members, particularly children, often resisted being monitored, creating a challenge for parents who had to balance supervision with respecting their children's independence. Despite efforts to prioritize consent and maintain open communication about monitoring, one-third of MDW participants acknowledged that their intentions to protect or monitor could sometimes feel invasive. MDW18 illustrated this tension: \textit{``It's my responsibility to help the child do well at school, and in return, children are expected to work hard to repay their parents for all their care and investment.''} This highlights how the Confucian value of filial piety, which emphasizes children's duty to honor and care for their parents, can create a hierarchical dynamic that sometimes blurs the line between care and control~\cite{naftali2010caged}.

\subsubsection{Legal protections and recommendations} \label{legal suggestions}


We found that most MDW participants had limited knowledge of privacy- and security-related laws (e.g., PIPL, CSL) and felt that existing legal frameworks, particularly labor laws, primarily protected employers' interests, overlooking the needs and rights of MDWs. They typically sought legal information only when conflicts arose, often learning through informal channels such as social media or word of mouth. However, this informal approach often resulted in receiving incomplete or inaccurate information, which worsened the challenges they faced in safeguarding their privacy and labor rights. For example, MDW21 argued: \textit{``This knowledge gap is superficial and lacks depth, leaving us without a comprehensive understanding of relevant laws. So workers may be vulnerable to privacy invasions or other legal issues because they are unaware of their rights beforehand [...]. Navigating the legal system is hard for us because we can't afford the cost of a lawsuit. These challenges prevent us from protecting our rights.''}

\textbf{Legal frameworks for disclosure and consent to monitoring.} Most MDW participants emphasized the importance of legal frameworks that required employers to disclose the presence, location, and purpose of surveillance devices, as well as obtain explicit consent. These laws should ensure that devices were used solely for security purposes, not to monitor personal activities. For example, MDW5 suggested: \textit{``To protect our privacy, efforts to minimize leaks should be increased. Future laws could require agencies or employers to disclose surveillance devices in homes and specify this in contracts. In addition, laws could limit the functionality or number of devices to prevent invasions, such as prohibiting spying while sleeping.''} Further, half of MDW participants emphasized the need for stringent regulations that forced companies of smart home devices to secure personal data and hold them responsible for breaches. MDW14 emphasized the importance of clear data handling practices and transparency from these companies: \textit{``We need laws to manage these companies. They should make sure these device owners should not be able to share footage with our private data without consent.''}


One-third of MDW participants highlighted the chaotic regulatory environment in China, pointing out the unclear responsibilities of agencies, employers, and MDWs. MDW25 said: \textit{``Although labor laws provide some protection, privacy and accountability aspects are not adequately covered. Agencies should be subject to stricter regulations''} Some MDW participants who identified as women emphasized the need for legal safeguards to ensure that employers treated them with respect and dignity, including protection against physical and verbal abuse. As MDW11 stated: \textit{``Protective measures should be in place to prevent sexual harassment of women workers by employers. Women are ashamed to speak out, especially to talk about this with employers or agencies. Without evidence, it is challenging for us to defend our rights [...] It's necessary to develop detailed laws to safeguard our privacy and physical safety in other people's homes.''}
\section{Discussion} \label{sec:discussion}
\subsection{Summary of Findings}

Our findings reveal that MDW participants prioritized safeguarding personal data and preventing physical or property-related harm. They expressed concerns about excessive data collection and perceived the required permissions for device functionality as intrusive. Surveillance cameras were the most concerning devices for participants. Although participants were uncertain about how to access privacy and security advice, they employed strategies to acquire relevant knowledge, including consulting hot-lines, referring to official government social media channels, and relying on word-of-mouth communication. In addition, MDW participants' privacy concerns varied significantly by gender, age, personality, and socioeconomic status. We found that social norms and cultural factors in China, such as collectivism, the normalization of public surveillance, and Confucian principles like filial piety, shaped MDW participants' perceptions of individual privacy \textbf{(RQ1, RQ2)}.

MDW participants encountered three key challenges: 1) employer surveillance through monitoring devices, reinforcing an unequal relationship and substantial power imbalance; 2) minimal agency support, including unclear contracts and insufficient training, which further increased MDW vulnerability and power disparities; and 3) limited capacity to advocate for their privacy rights due to a lack of legal knowledge and restricted ability to voice their concerns \textbf{(RQ2)}.

To address these challenges, MDW participants adopted coping strategies such as avoiding sensitive conversations in monitored areas. Some employers adjusted their surveillance practices to respect the privacy of MDW participants, for example, by turning off cameras when unnecessary. In some cases, MDW participants occasionally negotiated privacy measures directly with employers or through agencies. Participants pointed out the lack of legal protections for MDWs’ privacy and labor rights in China, stressing the need for regulations to oversee employer surveillance, require disclosure of monitoring devices in workplaces, regulate agency practices, and hold smart home device companies accountable \textbf{(RQ3)}.

Agencies, on the other hand, recognized MDWs' privacy and security needs and concerns; however, they often prioritized employers' demands over MDWs' privacy needs, such as strictly protecting clients' data rather than that of MDWs. Furthermore, although some agencies invited legal experts to discuss privacy and security, training typically focused on domestic work skills and legal knowledge, with limited attention to specific privacy issues \textbf{(RQ1)}.

\subsection{Contributions to Prior Work} \label{contributions}
In line with previous research, we found that gender dynamics shaped purchasing and usage preferences. In male-dominated households, most MDW participants and family members were women or female, and they were frequently viewed as ``passenger'' users~\cite{strengers2019protection,koshy2021we,kraemer2020further,geeng2019s,nicholls2020social,del2021controllable,pink2023smart,chambers2020domesticating} (see \textbf{\S\ref{mitigation practices and suggestions}}). While we explored how Chinese cultural factors, such as filial piety in Confucianism, influenced the privacy perceptions of MDWs and their families, our findings are consistent with previous research on general privacy concerns. They highlight the limited awareness of privacy issues among elderly family members, often leading these members to forfeit their privacy and control over personal data, subject to the equation of \textit{surveillance-as-care}~\cite{frik2019privacy,zou2024cross}. In some cases, MDW participants monitored their elderly family members in their own homes as part of their caregiving responsibilities. In this context, surveillance was not only deemed necessary to provide care and was seen as an expression of filial piety, but it was also influenced by the widespread normalization of surveillance as a public safety measure rather than as a privacy violation (see \textbf{\S\ref{suggestions to public}}).

Although previous studies have examined privacy concerns related to surveillance, highlighting the importance of keeping incidental users informed~\cite{chiang2024more,alladi2020consumer,meng2021owning,windl2023investigating,yao2019privacy}, we emphasize the importance of informing MDWs working in surveilled environments, highlighting the need for transparency in employment contracts to clearly define surveillance practices. Furthermore, while existing studies have examined power imbalances in employer-employee relationships due to surveillance, they have not explored the role of domestic worker agencies~\cite{bernd2022balancing,bernd2020bystanders,johnson2020beyond,ju2023re}. Our study addresses this gap by providing a nuanced understanding of how MDW-agency relationships influence MDWs' privacy and security (see \textbf{\S\ref{agency concerns}}). We also address this gap by offering a culturally nuanced perspective on how Chinese cultural and social norms influence MDWs' privacy concerns and behaviors (see \textbf{\S\ref{social view}}, \textbf{\S\ref{concerns in own home}}). Although Albayaydh and Flechais have examined the perspectives of agencies on MDW privacy needs and offered recommendations~\cite{albayaydh2024co}, our study focuses on the intricate power dynamics between MDWs, agencies, and employers, highlighting how surveillance intensifies power imbalances and the coping strategies that MDWs develop to navigate these challenges.

Although previous studies have examined the privacy concerns of Chinese device users~\cite{li2023will,liu2022privacy,huang2019perception} and the experiences of Filipino MDWs in Hong Kong and Macau~\cite{johnson2020beyond,ju2023re}, where the social and legal contexts differ significantly, our study examines the unique challenges faced by Chinese MDWs in mainland China. We explore how local policies, cultural norms (e.g., collectivism), and the involvement of domestic worker agencies shape the privacy and security experiences of MDWs. Specifically, while Yang et al. have examined job satisfaction in relation to labor control strategies (such as video monitoring and hongbao gifts), highlighting perceived discrimination as a mediating factor that explains the impact of these strategies on MDWs' job satisfaction, they have not addressed the unique privacy concerns and needs of Chinese MDWs in surveillance-intensive workplaces~\cite{yang2022act}. In contrast, our study focuses on the privacy and security concerns of MDWs in smart homes, examining the unique privacy risks posed by these technologies and providing actionable recommendations for stakeholders, such as policymakers, to mitigate these risks.

\subsection{Social and Cultural Factors Affecting Public Awareness} \label{suggestions to public}
\subsubsection{Impact of surveillance normalization and real-name system on privacy perceptions}
According to \textbf{\S \ref{concerns in workplace}} and \textbf{\S \ref{power dynamics}}, similar to previous studies~\cite{bernd2022balancing,johnson2020beyond}, our findings also indicate that the structure of surveillance systems fosters a panoptic effect~\cite{foucault2020panopticism}, causing MDWs to feel constantly watched, which in turn leads to noticeable changes in their behavior. Further, our findings align with those of Li et al.~\cite{li2023will}, who found that Chinese users often prioritize collective benefits over individual privacy. In a collectivist society that emphasizes public safety and social order, surveillance is frequently perceived as an essential tool rather than a violation of privacy. However, our study expands on this by demonstrating that, even within collectivist cultures, individuals have nuanced privacy concerns (see \textbf{\S\ref{perspectives on PSS in workplace}}). Unlike the resistance strategies employed by domestic workers in~\cite{ming2024wage}, our MDW participants rarely expressed active resistance to surveillance. They became more compliant, less spontaneous, and overly cautious, unsure of when they were being observed. The normalization of surveillance, driven by the widespread use of CCTV in public spaces, has significantly contributed to the acceptance of monitoring devices in China, further reinforcing the public's acceptance of being monitored~\cite{su2022explains,chin2022surveillance,chen2023maintainers}.

In addition, according to \textbf{\S\ref{social view}}, we point out how China's real-name system, which mandates individuals to register for services and platforms using their actual identities, has further diminished the public's sense of privacy~\cite{lee2016real,zhang2019large}. For many MDW participants with limited privacy awareness — accustomed to a society where surveillance is ubiquitous — the presence of monitoring devices at work was viewed merely as an extension of the public surveillance they encountered daily. This normalization made privacy invasions appear inevitable and even expected, diminishing the likelihood that they would challenge or question such practices in their workplaces. Consequently, constant surveillance induced feelings of anxiety and powerlessness, prompting these workers to adjust their behavior to align with perceived expectations, effectively self-regulating under the assumption that they were always observed, and often accepting to compromise their privacy in a state of resigned acceptance (see \textbf{\S\ref{concerns in workplace}} and \textbf{\S\ref{mitigation in workplace}}). Surveillance functioned as a control mechanism that reinforced power imbalances, benefiting employers while disadvantaging MDWs. This situation was further aggravated by the lack of transparent surveillance practices, placing MDWs in a vulnerable position. 

\subsubsection{Influence of social hierarchy and collectivism on privacy perceptions}
In contrast to WEIRD societies, where individual privacy is typically considered a fundamental right, privacy in China is often perceived within the framework of social responsibility and hierarchical relationships~\cite{hofstede1984culture,hofstede2011dimensionalizing}. Confucianism emphasizes social harmony, respect for hierarchy, and collective good over individual rights, which can lead to the perception that privacy is less important to ordinary people compared to the rich or famous~\cite{wang2015confucianism,wei2013confucian}. This difference may be attributed to the cultural emphasis on collectivism and respect for authority in China, which influences MDWs to accept monitoring as part of their employment conditions. Like other studies in Asian countries (see, e.g., ~\cite{sambasivan2018privacy}), we found that in a Confucian society, where social roles are clearly defined, the privacy of ``lower-level'' individuals, such as MDWs, is often considered less significant. This notion is further reinforced by the belief that individuals in higher social or economic positions deserve greater privacy protection, as their actions are perceived to have a more significant impact on society. As a result, many MDW participants perceived their privacy needs as secondary and did not actively pursue privacy and security protections, particularly in relation to workplace surveillance.

Furthermore, our findings revealed that MDW participants encountered a tension between their individual privacy rights and collective well-being, a conflict shaped by the collectivist values of Chinese society. While they might feel uncomfortable with or concerned about privacy violations, they often prioritized maintaining and fulfilling their roles within the hierarchical structure, especially as monitoring equipment in employers' homes has become a default setup in the domestic service industry (see \textbf{\S\ref{perspectives on PSS in workplace}}). This tendency aligns with the collectivist values prevalent in China, where the welfare of the group and social harmony take precedence over individual rights. It also reflects a broader dilemma, in which individuals hesitate to assert their personal rights due to concerns about disrupting group cohesion and the perceived low likelihood of achieving successful outcomes~\cite{steele2013pursuit,zhang2023individualism}. Many MDW participants were reluctant to raise privacy concerns on their own, fearing possible repercussions, such as strained relationships with employers or agencies, or believing that individual efforts would not lead to any significant change. Therefore, we emphasize that promoting collective action and aligning individual interests with broader group goals could be crucial steps in enhancing privacy protections for MDWs. However, starting such movements is difficult in the current context, where collectivism and respect for hierarchy are deeply embedded in the Chinese culture. Implementing policies that secure workers' privacy rights, alongside educating employers about ethical surveillance practices, could strike a balance between individual privacy needs and collective societal values. By presenting privacy protections as contributing to social harmony and the overall well-being of the community, such measures might be more readily accepted within a collectivist framework, thereby fostering gradual cultural shifts toward greater recognition of individual privacy concerns.

\subsubsection{Impact of Confucian values on privacy perceptions}
Our results in \textbf{\S\ref{definition of PSS}}, \textbf{\S\ref{chinese_cultural_and_social_factors}}, \textbf{\S\ref{mitigation in workplace}}, and \textbf{\S\ref{legal suggestions}} empirical evidence that the concept of ``privacy (yin si)'' is associated with derogatory connotations (e.g., shame)~\cite{mcdougall2004privacy,jingjing2023philosophy,wang2011protecting}, particularly for women~\cite{wang2020shameful}. Furthermore, we observed that many participants tended to hypothesize about how others might feel, such as men versus women, young versus old individuals, or poor versus wealthy people, rather than discussing their own feelings directly. This tendency (called ``to make remarks behind one’s back (yi lun)'') is a cultural communication style in China, where people may be less willing to discuss personal matters and are more likely to voice complaints about those in positions of authority who are perceived to have control over them~\cite{gao1998don}.

Culturally, domestic work in China is influenced by traditional views on gender and labor, particularly affecting MDWs, who are predominantly women. For instance, MDW participants felt uncomfortable negotiating with employers about issues related to physical privacy, viewing it as an ``unimaginable'' or inappropriate norm (see \textbf{\S\ref{legal suggestions}}). Moreover, the gender-based hierarchy rooted in Confucian values further shapes MDW attitudes toward privacy. Confucianism traditionally designates distinct roles for men and women, reinforcing a patriarchal structure where men are seen as protectors and providers, while women are expected to be cared for, often with an emphasis on submission and accommodation~\cite{rosenlee2023gender,lam1997cultural}. This gendered expectation can lead female MDWs to be more accepting of surveillance, as they may perceive it as a natural extension of their subordinate role within both the household and society. 

Additionally, we emphasize that the concept of filial piety, which highlights respect and duty toward one's parents and elders~\cite{ho1996filial}, plays a crucial role in shaping attitudes toward privacy. MDWs and their older family members, especially those from rural areas with deep Confucian influences, may prioritize their family's well-being and social harmony over their own personal privacy~\cite{madsen1984morality}. Older family members may perceive surveillance devices as acts of care from their children, interpreting them not as intrusions, but as expressions of filial responsibility. However, they may sometimes feel reluctant, driven by the pressure to conform to family expectations. Although filial piety highlights the importance of caring for one's parents, it does not inherently justify violating their privacy without consent. Similarly, younger children can also be monitored by their parents, such as observing their studies or online activities. However, our MDW participants did not perceive this as constant surveillance, but as a necessary parenting strategy. This perspective could be linked to the lasting influence of Confucian family values, which view children's academic success as part of parents' responsibilities~\cite{naftali2010caged,wu1996parental,liu2024cctv,shi2023monkey}. The Confucian emphasis on hierarchical relationships and obedience often makes it challenging for younger family members to voice objections or challenge authority~\cite{wang2022individual}. This dynamic underscores an inherent tension in Confucianism: its emphasis on familial harmony and care can inadvertently come at the expense of individual privacy rights and personal autonomy.

\subsection{Recommendations for Agencies} \label{suggestions to agencies}
Agencies act as intermediaries between MDWs and employers, but many MDW participants felt that agencies prioritized employers' needs over their well-being and rights, such as focusing on protecting employers' private information rather than addressing MDWs' privacy concerns (see \textbf{\S\ref{mitigation in workplace}}). They also reported that contracts often lacked details about surveillance practices and safety provisions, making them vulnerable to exploitation (see \textbf{\S\ref{agency concerns}}). Although the Ministry of Commerce issued a model contract for domestic services in 2014\footnote{\href{https://m.mofcom.gov.cn/article/h/redht/201405/20140500598286.shtml}{Notice of the Ministry of Commerce on Issuing the Model Household Service Contract}}, this official sample does not address workplace surveillance practices. Moreover, rapid technological advancements have outpaced existing laws, underscoring the urgent need for clearer communication and transparency regarding surveillance in employment contracts. Updates are necessary to align with emerging technologies, such as the integration of LLMs and AI into smart speakers, to ensure that privacy and security concerns of domestic workers are adequately addressed~\cite{seymour2023systematic}. We also found that existing MDW training programs focus more on professional skills than on privacy protection and legal knowledge (see \textbf{\S\ref{agency concerns}}, \textbf{\S\ref{legal suggestions}}). Hence, agencies should adopt more transparent and supportive practices, ensuring that the privacy and security needs of MDWs are considered alongside those of clients. This includes providing clear and understandable contracts that outline surveillance practices and safety provisions, and offering training that integrates tailored legal rights and privacy protections. Training should focus on practical, low-cost privacy protections that MDWs can easily employ to enhance their awareness of privacy and labor rights.

\subsection{Recommendations for Policymakers} \label{legal suggestion}
According to \textbf{\S\ref{agency concerns}} and \textbf{\S\ref{legal suggestions}}, participants pointed out the limitations and ambiguities in China's existing regulatory environment, particularly the absence of regulations governing domestic worker agencies. Our findings indicate that these agencies frequently collaborate with public authorities to access MDWs' private information but lack proper safeguards to protect their privacy (see \textbf{\S\ref{mitigation in workplace}}). 
Therefore, we stress the importance of establishing more transparent regulatory terms for agencies and employers, implementing clear and fair data handling practices, and ensuring accessible legal avenues for MDWs to seek redress in cases of privacy violations. These measures are crucial to protecting the private information of both MDWs and their clients. Specifically, the Ministry of Human Resources and Social Security (MOHRSS) should enhance its regulatory oversight of agencies, ensuring that these agencies comply with labor laws and regulations related to privacy and surveillance. In addition, safeguarding MDWs' privacy rights has become increasingly challenging due to ambiguous legal frameworks, including unclear provisions in PIPL and the absence of well-defined laws governing the domestic service industry (see \textbf{\S\ref{subsec:regulations}}). Therefore, we call for the establishment of clear and practical legal frameworks in collaboration with regulatory bodies, legal experts, and stakeholders in the domestic work market, considering the nuances of the Chinese regulatory context. The MOHRSS, in collaboration with local HR and social security offices, should ensure strict compliance with labor contracts and worker protections by establishing clearer standards and enforcing them consistently across different regions. Additionally, other government bodies, such as the Ministry of Civil Affairs and local governments, should play a role in regulating domestic workers' rights and protections, especially in cases where services intersect with social welfare programs. The All-China Federation of Trade Unions could also play a vital role in advocating for and protecting the rights of domestic workers, including their privacy. 

\section{Conclusion}
We conducted in-depth interviews with 26 MDWs and 5 domestic worker agencies in China, uncovering unique privacy and security challenges that Chinese MDWs face in multi-user smart homes. Our study first examines the influence of Chinese social and cultural norms (particularly Confucian values and collectivism) on shaping MDWs' perceptions and practices regarding privacy and security. We find that cultural factors, such as filial piety and gender-based hierarchies rooted in Confucianism, often lead MDWs and their families to compromise privacy, viewing surveillance as an expression of care rather than an invasion. Furthermore, we highlight how the widespread use of surveillance technologies intensifies power imbalances between MDWs, agencies, and employers, reinforcing a persistent sense of monitoring and control. Finally, we provide actionable recommendations for domestic worker agencies and policymakers, including integrating privacy education into training programs and establishing transparent communication and contractual agreements with regard to surveillance practices, particularly within China’s evolving and ambiguous regulatory landscape. By addressing gaps in existing literature, our study provides a culturally informed perspective on the privacy challenges faced by MDWs in non-Western smart home settings and suggests practical recommendations to improve the privacy and security of MDWs in Chinese smart homes.

\begin{acks}
We are grateful to our participants for their time and invaluable insights during the interviews. We also thank Chi Zhang and the anonymous reviewers for their valuable feedback. Shijing He is supported by a King’s-China Scholarship Council (K-CSC) PhD Scholarship, and Yaxiong Lei is supported by a Joint Scholarship from the University of St Andrews and China Scholarship Council.
\end{acks}


\bibliographystyle{ACM-Reference-Format}
\bibliography{main}

\end{CJK*}
\end{document}